\documentclass[aip,jcp,numerical,twocolumn, reprint]{revtex4-1}

\usepackage{graphicx}
\usepackage{url}
\usepackage[version=3]{mhchem}
\usepackage{siunitx}
\usepackage{amsmath,amssymb}
\usepackage{epsfig}
\usepackage[utf8]{inputenc}

\begin{document}
\setlength{\parskip}{0.3\baselineskip}

\title{Crystallization mechanism in melts of short n-alkane chains}

\author{Muhammad Anwar}
\author{Francesco Turci}

\author{Tanja Schilling} 
\affiliation{Universit\'e du Luxembourg, Theory of Soft Condensed Matter, L-1511 Luxembourg, Luxembourg}


\date{\today}

\begin{abstract}
We study crystallization in a model system for eicosane (C20) by means of molecular dynamics simulation and we identify the microscopic mechanisms of homogeneous crystal nucleation and growth. For the nucleation process, we observe that chains first align and then straighten. Then the local density increases and finally the monomer units become ordered positionally. The subsequent crystal growth process is characterized by a \textit{sliding-in} motion of the chains. Chains preferably attach to the crystalline cluster with one end and then move along the stems of already crystallized chains towards their final position. This process is cooperative, i.e.~neighboring chains tend to get attached in clusters rather than independently.
\end{abstract}

\pacs{}

\maketitle


Despite the long-standing research interest in crystallization in polymer melts, many fundamental aspects of the crystal nucleation and growth mechanisms are still subject of discussion \cite{Strobl2009}.
Experiments in this field are usually restricted to a spatial and temporal resolution that is too coarse to capture atomistic details of individual nucleation events. Thus molecular dynamics provides an ideal instrument to complement experiment and offer insight into the mechanisms on the atomistic scale.

Given the high degree of complexity that long polymer chains pose both, from the conceptual and the numerical point of view (due to folding and entanglement), a basic comprehension of how even relatively short chains crystallize is of fundamental importance in order to build a coherent theory.
Crystal nucleation in alkanes has been addressed in several computer simulation studies in the 90s \cite{esselink:9033,takeuchi:5614,PhysRevLett.80.991,fujiwara:9757} and a scenario for the nucleation mechanism has been suggested. Due to the limited computer resources available at the time, however, these works were based on one simulation trajectory each (with the exception of ref. \cite{esselink:9033}). 
The first direct computations of homogeneous nucleation rates in n-alkanes by means of computer simulation have been presented by Rutledge and co-workers in the past years \citep{Yi:2009im,Yi:2011kd}. These studies were focussed on the nucleation and growth rates and the free energy landscape associated with the crystallization process rather than the microscopic mechanisms.
Very recently, also simulation results on nucleation rates  \citep{doi:10.1021/ma4004659} and growth mechanisms \cite{yamamoto:054903,doi:10.1021/ma102380m} in systems of chains longer than the entanglement length have been presented.
 
Considering the limited amount of data available in the literature on the nucleation and growth mechanism in short chain alkanes, we have revisited the problem and present here a detailed analysis of the formation of crystal nuclei from the homogeneous melt and the subsequent growth process.

\section{Simulation method and parameters}
We use a standard united atom model for polyethylene\citep{Paul:1995uq} in which point-like particles represent \ce{CH_{2}} and \ce{CH_{3}} groups. Non-bonded particles interact via Lennard-Jones potentials and particles bonded along the chain interact via harmonic bond length and bond angle potentials as well as a dihedral potential. We set the system parameters as in ref.~\citep{Waheed:2005km}, with the exception of the Lennard-Jones cutoff radius which we set to $r_{c}^{LJ}=2.5 \sigma$. For the system parameters chosen, the Lennard Jones radius $\sigma$ corresponds to $\SI{4.01}{\angstrom}$. We express all physical quantities in units of the intrinsic units of the Lennard Jones model (i.e.~the particle mass $m$, the interaction energy $\epsilon$ and resulting timescale $\tau = \sqrt{m \sigma^2/\epsilon}$), apart from the temperatures and pressures which we converted to Kelvin resp.~atmospheres. 

We used the ESPResSo package \citep{Limbach:2006cj} with an \textit{ad hoc} implementation of the dihedral potential in order to  perform molecular dynamics simulations. The simulations were carried out in the NPT ensemble (constant number of chains, pressure, temperature) by means of a Langevin thermostat with a friction coefficient $\gamma=0.5 1/\tau$ , and an Andersen based barostat. The barostat reduced the cubic box linear dimension during the crystallization process in order to keep a pressure of 1 atm, with a piston mass  $M=10^{-5}m$\citep{Kolb:1999vt}. The integration time step used throughout the simulations is $dt=0.005\, \tau$. 
To initiate the simulations, we equilibrated 500 chains of 20 particles each  at a temperature $T=400 K$ (well above the melting temperature $T_{m}=310 K$ \citep{Yi:2011kd}). 25 independent configurations were then quenched to $T=250 K$, where we observed homogeneous crystal nucleation and growth. 

\begin{figure}[t]
\begin{center}
\includegraphics[width=0.45\textwidth]{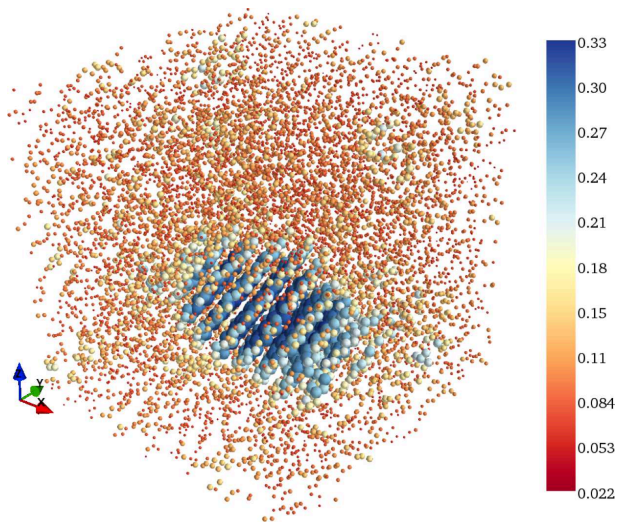}
\caption{(color online) Use of averaged local bond order parameters in order to determine the crystalline structures. The particles in this simulation snapshot are colour-coded according to $\bar{q}_{6}$ } \label{composition1}
\end{center}
\end{figure}

\textit{Order parameters - }
In order to distinguish the crystalline from the fluid-like regions of 
the system, we define several order parameters:

\begin{itemize}
\item 
The local density is measured by means of Voronoi tesselation, i.e.~the density at the position of particle $i$ is defined as the inverse of the volume of particle $i$'s voronoi cell.

\item 
We measure the global alignment of chains in terms of the nematic order 
parameter S, which is the largest eigenvalue of 
\[
Q_{\alpha \beta} = \frac{1}{N_{\rm cn}}\sum_{i=1}^{N_{\rm cn}}\left(\frac{3}{2} \hat{u}_{j\alpha} \hat{u}_{j\beta} - \frac{1}{2} \delta_{\alpha\beta}\right) \quad ,
\]
where $N_{\rm cn}$ is the number of chains for which the calculation is performed, $\hat{u}_j$ is the unit vector parallel to the end-to-end vector of chain $j$, $\delta$ is the Kronecker delta and $\alpha, \beta = x,y,z$. \cite{P.G.deGennes1995}. 

\item
To monitor the local alignment of segments of chains, we identify for a 
given particle 
$i$ the neighbouring particles $j$ (i.e.~the particles that lie within 
a distance $r_{c}=1.4\sigma$ from particle $i$). For every neighbour $j$ 
we determine
\begin{equation}
\theta_{ij}=\arccos(\hat{e}_{i}\cdot\hat{e}_{j})
	\begin{cases}
	\leq 10\si{\degree} \mbox{ ``aligned''}\\
	>10\si{\degree} \mbox{ ``non-aligned''}
	\end{cases}
\label{alignment}
\end{equation}
where $\hat{e}_{j}$ are unit vectors pointing from the position of particle $j-1$ to the position of particle $j+1$ in a given chain. 
Particles that have at least 13 ``aligned'' neighbours are called crystalline. We obtained this threshold number from an analysis of the probability distributions of aligned neighbors in the bulk melt and the bulk crystal. It distinguishes melt-like configurations from crystals.

\item
In order to identify crystalline clusters, we use a standard clustering algorithm. This proceeds by picking a particle and checking whether it is crystalline. If so, we count it as the first particle of a cluster and analyze its shell of neighbours, including into the cluster neighboring particles that are also crystalline. In this way, we move recursively from neighbour to neighbour to detect the complete cluster and compute its size. If no new crystalline neighbour is found, the cluster is complete and we proceed with the other particles of the system to detect further clusters, if there are any.

\item
To characterize crystal order in terms of particle positions rather than segment alignment, we use \textit{local bond orientational order} parameters. (The term ``bond order'', which is commonly used for this type of parameter in the context of monatomic systems, might be misleading in the context of polymers. It refers to the orientation of the vector between any pair of neighbouring particles, not just to bonds along the chain.)
Bond orientation parameters characterize the local positional structure by projection of the positions of a particle's neighbours onto spherical harmonics. Rather than the original definition by Steinhardt \citep{1983PhRvB..28..784S} we use a recent extension \citep{Lechner:2008vz} which exploits additional information derived from the second shell of neighbours, defining the so called \textit{averaged local bond order parameters} (ALBO). This definition requires the computation of the complex vector $q_{l}(i)$
\begin{equation}
q_{lm}(i)=\frac{1}{N_{b}(i)}\sum_{j=1}^{N_{b}(i)}Y_{lm}(\vec{r}_{ij}) \end{equation}
where $N_{b}(i)$ corresponds to the number of nearest neighbors of particle $i$ and $Y_{lm}(\vec{r}_{ij})$ are the spherical harmonics. Averaging over the neighbors of particle $i$ and particle $i$ itself
\begin{equation}
\bar{q}_{lm}(i)=\frac{1}{\tilde{N}_{b}(i)}\sum_{k=0}^{\tilde{N}_{b}(i)}q_{lm}(k),
\end{equation}
and summing over all the harmonics
\begin{equation}
\bar{q}_{l}(i)=\sqrt{\frac{4\pi}{2l+1}\sum_{m=-l}^{l}|\bar{q}_{lm}(i)|^{2}}
\end{equation}
one gets the final value of the locally averaged bond order parameter $\bar{q}_{l}$. Fig.~\ref{composition1} shows a system snapshot labelled according to $\bar{q}_{6}$. The crystallite embedded in the melt is clearly visible.

We tested the ALBO parameters determining the neighbours both according to the spherical cutoff at $r_c=1.4\sigma$ as well as using the neighbor list resulting from the Voronoi \citep{Rycroft:2009ff} tessellation. We observe only negligible quantitative differences between the two approaches, justifying the choice of the cutoff radius.
\end{itemize}

During the simulation runs, we used the size of the largest cluster of crystalline particles (crystalline according to the local alignment parameter) as the main reaction coordinate to track the formation of crystalline nuclei and their growth.

\section{Nucleus formation}
To identify crystal nucleation events, we perform a \textit{committor} analysis \citep{2002ARPC...53..291B}: we determine $p_{\rm crystal}(n_c)$, the probability that a trajectory initiated from a given cluster size $n_c$ ends in a stable crystalline state. The cluster size for which $p_{\rm crystal }(n_c)=0.5$ is the typical size of the \textit{critical nucleus}.

The analysis has been performed considering 7 different cluster sizes ranging from 30 to 200 monomer units. For each of these three independent configurations were extracted out of the 25 independent runs. We randomized the velocities of these configurations eight times, and thus generated 24 new trajectories per cluster size, which were run until either a stable crystal or a melt (cluster size $<30$) configuration was reached.
 This type of analysis has the advantage that it is based on the kinetics of the transformation process only and does not require an underlying free-energy landscape model, such as e.g.~an analysis in terms of classical nucleation theory. We find that the critical nucleus has a size of $80\pm20$ particles (i.e.~repeat units). The uncertainty is mainly due to our choice of crystallinity parameter as the main reaction coordinate to interpret the committor analysis. This shows that additional parameters are needed to properly capture the dynamics of the crystallization process.

\begin{figure}[t]
\centering
\includegraphics[width=0.5\textwidth]{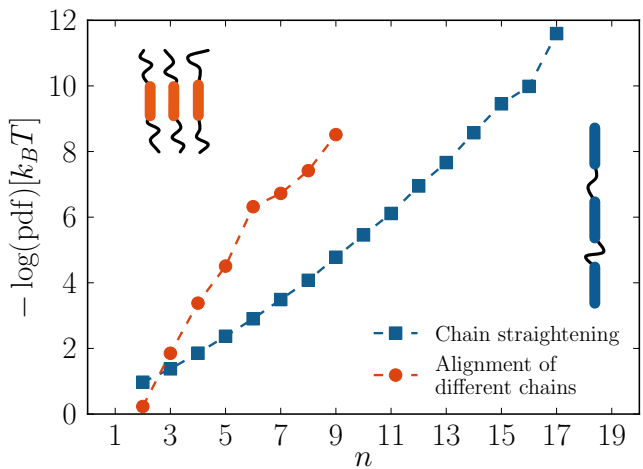}
\caption{(color online) Effective free energies associated with straightening of individual chains (blue squares) and alignment of neighbouring chains (red circles) as a function of the size of clusters of aligned segments $n$. The colored segments in the sketches represent the selection criteria used for the computation of the corresponding probabilities.  
}Yi:2009im
\label{free}
\end{figure}

\begin{figure}[t]
\centering
\includegraphics[width=\columnwidth]{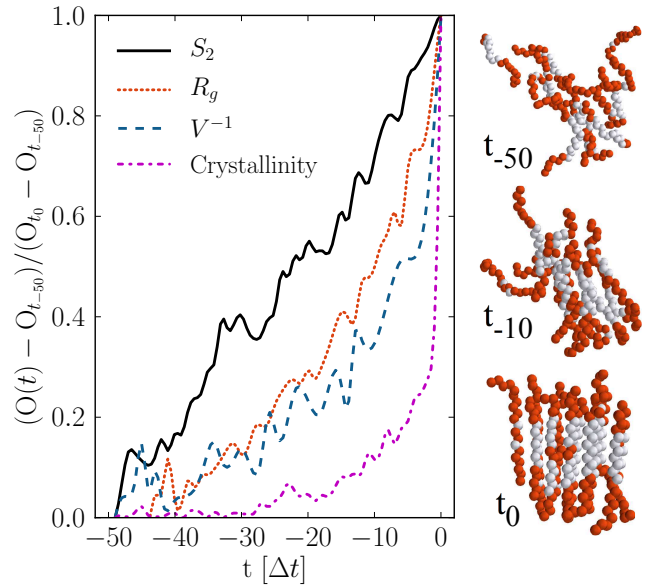}
\caption{(color online) Relative variation of several observables (O) from the melt to the formation of a critical nucleus for the particles involved in the nucleus: the nematic order parameter $S_{2}$ (black, continuous), the radius of gyration $R_{g}$ (red, dotted), the inverse of the Voronoi cell volume $V$ (blue, dashed) and the crystallinity order parameter (purple, dash-dotted) corresponding to the largest cluster size. The curves are averaged over 25 independent trajectories progressing backward in time from the nucleation time $t=t_{0}$ in steps $\Delta t$ to $t=-50\Delta t$. On the right side, we present three snapshots of the nucleus chains. The particles that form the nucleus at time $t_{0}$ are highlighted in grey. The chains are already initially prolate and undergo orientational ordering before they straighten further. Finally a cluster of aligned, hexagonally placed chains is formed.} 
\label{precursor}
\end{figure}

\begin{figure}[h!tbp]
\centering
\includegraphics[width=\columnwidth]{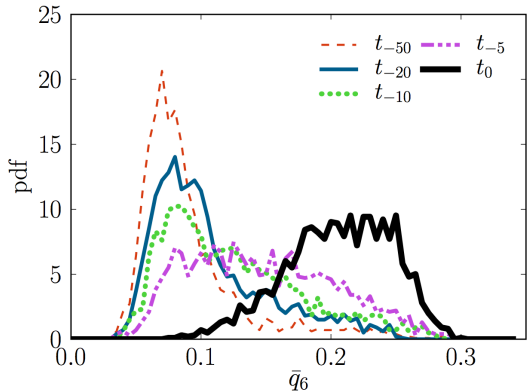}
\caption{(color online) Probability distributions of the averaged local bond order parameter $\bar{q}_{6}$ computed at different times for those particles that form the crystallite at $t_{0}$. 
}
\label{q6dists}
\end{figure}

To form a nucleus, a critical number of segments of neighbouring chains 
need to align (see fig.~\ref{free}). The circles (red 
online) in fig.~\ref{free} show the free 
energy change in the metastable melt associated with 
the occurrence of such a configuration, i.e.~with the occurrence of a cluster of aligned neighbouring segments that belong to $n$ different chains.  
(A segment is defined for a monomer unit $i$ as the vector connecting the center of monomer $i-1$ to the center of monomer $i+1$.) 
For comparison we have plotted the free energy change associated with straightening individual chains (squares), i.e.~with finding $n$ aligned segments within the same chain. 
The relatively low free energie changes reflect the long persistence length of polyethylene of approximately 8 units \citep{Yi:2009im}).

Locally aligned clusters containing segments of more than 9 chains are extremely unlikely to appear by spontaneous fluctuation. In contrast, the melt displays a non-negligible probability to find piece-wise straightened chains, where up to 14-15 out of 20 segments can point in the same direction. Forming a locally ordered (aligned) environment is therefore much more expensive in terms of free energy than straightening individual chains. Similar observations have been made by Takeuchi \cite{takeuchi:5614} and Miura and co-workers \cite{PhysRevE.63.061807}, who concluded that the nucleation process was initiated by chain straightening and then completed by chain orientation and crystallization. We will show in the following, that this conclusion is not completely correct. In order to further determine which conditions in the melt structure favour crystallization, we identify the particles that form a critical nucleus and analyze their previous pathway in time. We name $t_{0}$ the time at which a crystalline cluster of about 80 particles is formed. We then proceed backwards in time in 
steps of $\Delta t=\tau_{D}/20$, where $\tau_{D}=4 \cdot 10^{5}dt$ is the center of mass diffusion time in the supercooled melt. At
$-50\Delta t$ all the particles that belonged 
to the nucleus at $t_{0}$ are indistinguishable from the ones of the melt 
according to their structural and orientational properties.  
We analyze 25 independent trajectories in terms of the average radius of gyration $R_{g}$ of all chains that are part of the nucleus at $t_0$, the global alignment S of these chains, the average volume $V$ of the Voronoi\citep{Rycroft:2009ff} cell associated to each particle that is part of the nucleus, its crystallinity parameter and the average local bond order parameter $\bar{q}_{6}$. 
In Fig.~\ref{precursor} we show the relative variations of these quantities with respect to the values they had $-50\Delta t$. For $\bar{q}_{6}$ we show the evolution of the entire distribution rather than just the average (fig.~\ref{q6dists}), because the average is still dominated by the peak at liquid-like $\bar{q}_{6}$ at times when there is already a clearly discernible shoulder at crystalline $\bar{q}_{6}$.
Approaching the formation of the critical nucleus at $t_0$, we observe first an increase in the global orientational order S, then an increase in the radius of gyration and in the local density, and finally local positional and orientational order are established. We conclude that already in the melt the chains are sufficiently prolate to undergo an ordering transition similar to the isotropic-nematic transition 
in liquid crystals. Only once they have formed an oriented aggregate, they start straightening. (This observation stands in contrast to what has been suggested in earlier work  \citep{takeuchi:5614,PhysRevE.63.061807}, but is similar to recent results of Luo and Sommer\citep{doi:10.1021/ma102380m}).
\begin{figure}[t]
\centering
\includegraphics[width=\columnwidth]{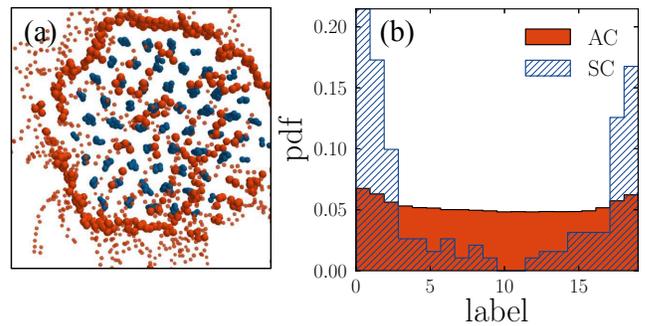}
\caption{(color online)  (a) Top view of a cluster. Crystalline particles (blue) and surface chains (red). Note the hexagonal arrangement of the chains and the relatively low coverage of the top surface by surface chains. (b) Normalized histograms of the surface particles versus label of a particle in the chain (0 to 19): all surface particles (filled histogram) and only those that belong to chains successfully attached after $\tau_{D}$ (dashed histogram).}
\label{composition2}
\end{figure}
\begin{figure*}[!th]
\centering
\includegraphics[width=\textwidth]{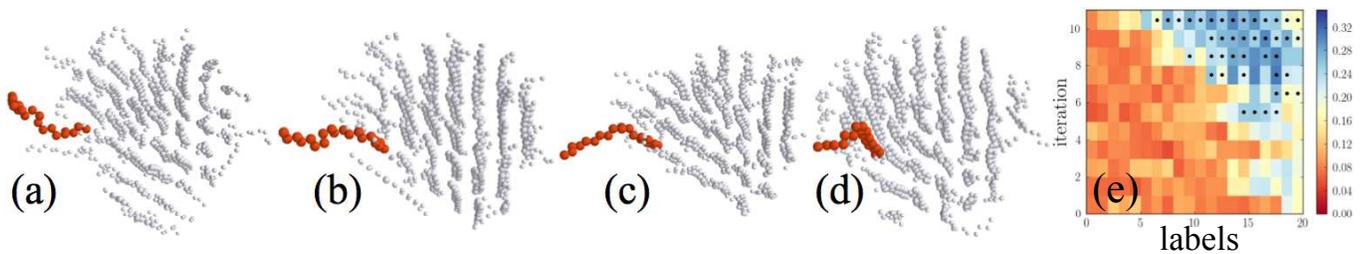}
\caption{(color online) Trajectory of a chain attaching via the sliding process: (a-d) Snapshots of the time evolution, with big red beads representing the attaching chain, medium sized gray beads being the cluster of crystalline particles and small gray beads being particles that belong to the cluster chains but are not crystalline; (e) Time evolution of the $\bar{q}_{6}$ order parameter for every particle in the chain, with black dots highlighting particles that are identified as crystalline according to the alignment criterium (eq. \ref{alignment}). Every iteration corresponds to a single $\Delta t = \tau_{D}/20$. }
\label{composition3}
\end{figure*}
\begin{figure}[hbt]
\centering
\includegraphics[width=\columnwidth]{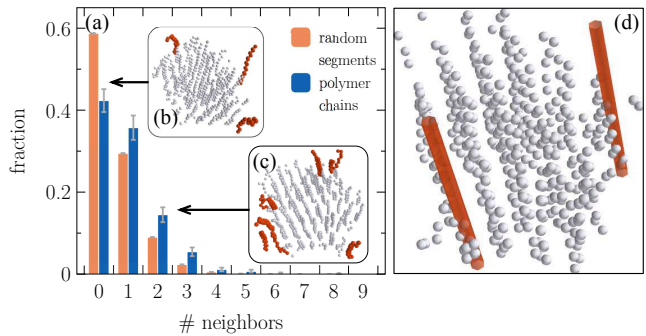}
\caption{(color online)(a) Correlation between attachment events: The dark (blue) bars represent the distribution of neighbouring attachment events 
as resulting from the analysis of 30 growth trajectories of length $30\Delta t$.
The light (red) color bars represent the distribution of neighbouring attachment events for a Monte-Carlo sampling of non-interacting cylinders attached at random sites picked uniformly on the surface of the crystal. Simultaneous attachment of neighbouring chains is more likely to occur in the interacting system than in the non-interacting system. (For detailed definition of terms, please see main text.) Isolated (b) and multiple (c) attachment events are shown in the insets. (d) Schematic illustration of a configuration of random segments placed at the surface of the crystalline cluster. Their direction corresponds to the average direction of the end-to-end vectors of the cluster chains.  
}
\label{composition4}
\end{figure}

\section{Pathway to growth}
Once a stable nucleus is formed, crystal growth proceeds via the successive attachment of new chains and a lamellar structure develops. For our choice of parameters, which corresponds to 19\% supercooling, we measure a growth rate of
approximately 25 particles $/\tau_{D}$.

To extract the attachment mechanism, we now consider only those pieces of 
simulation trajectories in which a cluster grows from a size of 300 particles 
to 900 
particles. This restriction serves to reduce any bias due to system size 
effects or merging of different clusters. Along every piece of trajectory, 
we take 
configuration snapshots at time intervals $\Delta t$. In each 
snapshot, we identify the crystallite and the ``surface chains'', i.e.~chains 
that are not part of the cluster but contain at least one particle with a 
distance of less than $1.4\sigma$ from the cluster.
Fig.~\ref{composition2} shows a typical system snapshot, in which all 
crystalline particles are labelled in blue and surface chains are labelled 
in red.

Now we ask what distinguishes those surface chains that will be attached 
to the cluster from those that will not be attached. We define 
an ``attaching chain'' as a surface chain of which at least 
seven repeat units will be part of the cluster within the next $\tau_D$. 
The choice of this threshold value is based on our empirical observation that once a chain is attached to the cluster with more than seven particles it does not detach anymore.

Neither the distributions of the radii of gyration nor of the $\bar{q}_6$ 
values differ between attaching and non-attaching chains (data not shown). 
But there 
is a clear difference in the position of the ends of the chain with respect to 
the cluster. Fig.~\ref{composition3} shows the distribution of 
particle labels (indicating the position of a particle along the chain) of 
those particles that are closer 
than $1.4\sigma$ to the cluster when the surface chain is identified, 
plotted for all surface 
chains (dashed histogram) and for the attaching chains only 
(filled histogram). Clearly, surface chains that have an end
close to the 
cluster are far more likely to be attached than surface chains that have the 
middle close to the cluster. This suggests that the initial stages of the 
attachment process are driven by the motion of the most mobile chain segments 
and that the crystallization of new chains is initiated at the ends. 
This mechanism is specific to short chains and it stands 
in contrast to folded chain crystallization for longer 
chains\cite{yamamoto:054903,doi:10.1021/ma102380m}.

With this picture in mind, we investigate how the remaining parts of a chain are attached to the cluster. Every $\Delta t$ we plot $\bar{q}_{6}$ for each 
particle in an attaching chain. This gives us a 
``particle label versus time map'' 
for each attaching chain (see Fig.~\ref{composition3}(e)). Based on visual 
inspection, we grouped these maps into classes and 
then compared representative maps for each class with movies of the 
corresponding molecular 
dynamics trajectories. As the predominant attachment mechanism we identify a \textit{sliding-in} motion: a particle located at one end of a chain is attached on the lateral surface of the cluster; this contributes to the increase of the local order of the neighboring particles, which enter the cluster region one by one in a single file. Typically (see Fig.~\ref{composition3}(a-d)) the increase of local order is accompanied by an increase of the radius of gyration, so that the chains are piece-wise straightened, often assuming L-shaped conformations. The end monomer that is attached first moves along the direction given by the nearest cluster chains, and guides the attachment of the rest of chain. 

Yet, the attachment process is not simply characterized by the stochastic motion of single chains in the melt that randomly attach to the cluster in an uncorrelated manner. It is a correlated process, as demonstrated in Fig.~\ref{composition4}. Between all chains that were attached in a given frame, we computed the pairwise distances (where the distance between two chains was defined as the distance between the closest pair of particles of the chains). If the distance was less than $1.4\sigma$ we considered two chains as ``neighbours''. Fig.~\ref{composition4} shows the frequency with which clusters of neighbouring chains have been attached. 

To compare this distribution with that of an uncorrelated process, we sampled 
the attachment statistics of non-interacting cylinders on the surface of 
the clusters. For each cluster configuration, we 
picked random sites uniformly distributed on the surface of the crystal and 
placed cylinders at these sites 
(see Fig.~\ref{composition4} (d)). The cylinders were 
oriented parallel to the average orientation of the chains in the crystallite. 
Their ``contact site'' with the cluster was picked
uniformly distributed along their length. For each crystallite, we picked as many cylinders as attaching chains 
had been observed, and produced 1000 different realizations of attachment 
events. Then we averaged the results over all crystallites.
Fig.~\ref{composition4} shows that about the 58\% of 
the attaching chains in the interacting system are in contact with at least 
one other attaching chain, while only the 41\% of the non-interacting 
cylinders on the same crystallite surface are. Snapshots of isolated (b) and 
multiple (c) attachment events are shown in the insets.

\section{Conclusions}
We have analyzed the formation of the critical nucleus and the crystal growth process in a model system of eicosane. We have determined via committor analysis the characteristic nucleus size and we have shown that the chains that form the critical nucleus first align, then straighten, and finally the local crystal structure forms. The growth of the crystal advances mainly through a \textit{sliding-in} process on the lateral surface, which takes place in a correlated way, i.e. chains tend to get attached in clusters.

\begin{acknowledgments}
We thank Jens-Uwe Sommer, Roland Sanctuary, J\"org Baller and 
Carlo Di Giambattista for stimulating discussions. 
This project has been financially supported by the National Research Fund (FNR) within the CORE project Polyshear and the INTER-DFG project Crystallization. 
Computer simulations presented in this paper were carried out using the HPC facility of the University of Luxembourg.
\end{acknowledgments}
%


\begin{thebibliography}{10}%
\makeatletter
\providecommand \@ifxundefined [1]{%
 \ifx #1\undefined \expandafter \@firstoftwo
 \else \expandafter \@secondoftwo
\fi
}%
\providecommand \@ifnum [1]{%
 \ifnum #1\expandafter \@firstoftwo
 \else \expandafter \@secondoftwo
\fi
}%
\providecommand \enquote [1]{``#1''}%
\providecommand \bibnamefont  [1]{#1}%
\providecommand \bibfnamefont [1]{#1}%
\providecommand \citenamefont [1]{#1}%
\providecommand\href[0]{\@sanitize\@href}%
\providecommand\@href[1]{\endgroup\@@startlink{#1}\endgroup\@@href}%
\providecommand\@@href[1]{#1\@@endlink}%
\providecommand \@sanitize [0]{\begingroup\catcode`\&12\catcode`\#12\relax}%
\@ifxundefined \pdfoutput {\@firstoftwo}{%
 \@ifnum{\z@=\pdfoutput}{\@firstoftwo}{\@secondoftwo}%
}{%
 \providecommand\@@startlink[1]{\leavevmode}%
 \providecommand\@@endlink[0]{}%
}{%
 \providecommand\@@startlink[1]{%
  \leavevmode
  \pdfstartlink
   attr{/Border[0 0 1 ]/H/I/C[0 1 1]}%
   user{/Subtype/Link/A<</Type/Action/S/URI/URI(#1)>>}%
  \relax
 }%
 \providecommand\@@endlink[0]{\pdfendlink}%
}%
\providecommand \url  [0]{\begingroup\@sanitize \@url }%
\providecommand \@url [1]{\endgroup\@href {#1}{\urlprefix}}%
\providecommand \urlprefix [0]{URL }%
\providecommand \Eprint[0]{\href }%
\@ifxundefined \urlstyle {%
  \providecommand \doi [1]{doi:\discretionary{}{}{}#1}%
}{%
  \providecommand \doi [0]{doi:\discretionary{}{}{}\begingroup
  \urlstyle{rm}\Url }%
}%
\providecommand \doibase [0]{http://dx.doi.org/}%
\providecommand \Doi[1]{\href{\doibase#1}}%
\providecommand \selectlanguage [0]{\@gobble}%
\providecommand \bibinfo [0]{\@secondoftwo}%
\providecommand \bibfield [0]{\@secondoftwo}%
\providecommand \translation [1]{[#1]}%
\providecommand \BibitemOpen[0]{}%
\providecommand \bibitemStop [0]{}%
\providecommand \bibitemNoStop [0]{.\EOS\space}%
\providecommand \EOS [0]{\spacefactor3000\relax}%
\providecommand \BibitemShut [1]{\csname bibitem#1\endcsname}%
\bibitem{Strobl2009}%
  \BibitemOpen
  \bibfield{author}{%
  \bibinfo {author} {\bibfnamefont{G.}~\bibnamefont{Strobl}},\ }%
  \bibfield{journal}{%
  \Doi{10.1103/RevModPhys.81.1287}{\bibinfo {journal} {Rev. Mod. Phys.}}\ }%
  \textbf{\bibinfo {volume} {81}},\ \bibinfo {pages} {1287} (\bibinfo {month}
  {Sep}\ \bibinfo {year} {2009})\BibitemShut{NoStop}%
\bibitem{esselink:9033}%
  \BibitemOpen
  \bibfield{author}{%
  \bibinfo {author} {\bibfnamefont{K.}~\bibnamefont{Esselink}}, \bibinfo
  {author} {\bibfnamefont{P.~A.~J.}\ \bibnamefont{Hilbers}},\ and\ \bibinfo
  {author} {\bibfnamefont{B.~W.~H.}\ \bibnamefont{van Beest}},\ }%
  \bibfield{journal}{%
  \Doi{10.1063/1.468031}{\bibinfo {journal} {The Journal of Chemical Physics}}\
  }%
  \textbf{\bibinfo {volume} {101}},\ \bibinfo {pages} {9033} (\bibinfo {year}
  {1994})\BibitemShut{NoStop}%
\bibitem{takeuchi:5614}%
  \BibitemOpen
  \bibfield{author}{%
  \bibinfo {author} {\bibfnamefont{H.}~\bibnamefont{Takeuchi}},\ }%
  \bibfield{journal}{%
  \Doi{10.1063/1.477179}{\bibinfo {journal} {The Journal of Chemical Physics}}\
  }%
  \textbf{\bibinfo {volume} {109}},\ \bibinfo {pages} {5614} (\bibinfo {year}
  {1998})\BibitemShut{NoStop}%
\bibitem{PhysRevLett.80.991}%
  \BibitemOpen
  \bibfield{author}{%
  \bibinfo {author} {\bibfnamefont{S.}~\bibnamefont{Fujiwara}}\ and\ \bibinfo
  {author} {\bibfnamefont{T.}~\bibnamefont{Sato}},\ }%
  \bibfield{journal}{%
  \Doi{10.1103/PhysRevLett.80.991}{\bibinfo {journal} {Phys. Rev. Lett.}}\ }%
  \textbf{\bibinfo {volume} {80}},\ \bibinfo {pages} {991} (\bibinfo {month}
  {Feb}\ \bibinfo {year} {1998})\BibitemShut{NoStop}%
\bibitem{fujiwara:9757}%
  \BibitemOpen
  \bibfield{author}{%
  \bibinfo {author} {\bibfnamefont{S.}~\bibnamefont{Fujiwara}}\ and\ \bibinfo
  {author} {\bibfnamefont{T.}~\bibnamefont{Sato}},\ }%
  \bibfield{journal}{%
  \Doi{10.1063/1.478941}{\bibinfo {journal} {The Journal of Chemical Physics}}\
  }%
  \textbf{\bibinfo {volume} {110}},\ \bibinfo {pages} {9757} (\bibinfo {year}
  {1999})\BibitemShut{NoStop}%
\bibitem{Yi:2009im}%
  \BibitemOpen
  \bibfield{author}{%
  \bibinfo {author} {\bibfnamefont{P.}~\bibnamefont{Yi}}\ and\ \bibinfo
  {author} {\bibfnamefont{G.~C.}\ \bibnamefont{Rutledge}},\ }%
  \bibfield{journal}{%
  \bibinfo {journal} {Journal of Chemical Physics}\ }%
  \textbf{\bibinfo {volume} {131}},\ \bibinfo {pages} {134902} (\bibinfo {year}
  {2009})\BibitemShut{NoStop}%
\bibitem{Yi:2011kd}%
  \BibitemOpen
  \bibfield{author}{%
  \bibinfo {author} {\bibfnamefont{P.}~\bibnamefont{Yi}}\ and\ \bibinfo
  {author} {\bibfnamefont{G.~C.}\ \bibnamefont{Rutledge}},\ }%
  \bibfield{journal}{%
  \bibinfo {journal} {Journal of Chemical Physics}\ }%
  \textbf{\bibinfo {volume} {135}},\ \bibinfo {pages} {024903} (\bibinfo {year}
  {2011})\BibitemShut{NoStop}%
\bibitem{doi:10.1021/ma4004659}%
  \BibitemOpen
  \bibfield{author}{%
  \bibinfo {author} {\bibfnamefont{P.}~\bibnamefont{Yi}}, \bibinfo {author}
  {\bibfnamefont{C.~R.}\ \bibnamefont{Locker}},\ and\ \bibinfo {author}
  {\bibfnamefont{G.~C.}\ \bibnamefont{Rutledge}},\ }%
  \bibfield{journal}{%
  \bibinfo {journal} {Macromolecules}\ }%
  \textbf{\bibinfo {volume} {46}},\ \bibinfo {pages} {4723} (\bibinfo {year}
  {2013})\BibitemShut{NoStop}%
\bibitem{yamamoto:054903}%
  \BibitemOpen
  \bibfield{author}{%
  \bibinfo {author} {\bibfnamefont{T.}~\bibnamefont{Yamamoto}},\ }%
  \bibfield{journal}{%
  \Doi{10.1063/1.4816707}{\bibinfo {journal} {The Journal of Chemical
  Physics}}\ }%
  \textbf{\bibinfo {volume} {139}},\ \bibinfo {eid} {054903} (\bibinfo {year}
  {2013})\BibitemShut{NoStop}%
\bibitem{doi:10.1021/ma102380m}%
  \BibitemOpen
  \bibfield{author}{%
  \bibinfo {author} {\bibfnamefont{C.}~\bibnamefont{Luo}}\ and\ \bibinfo
  {author} {\bibfnamefont{J.-U.}\ \bibnamefont{Sommer}},\ }%
  \bibfield{journal}{%
  \Doi{10.1021/ma102380m}{\bibinfo {journal} {Macromolecules}}\ }%
  \textbf{\bibinfo {volume} {44}},\ \bibinfo {pages} {1523} (\bibinfo {year}
  {2011})\BibitemShut{NoStop}%
\bibitem{Paul:1995uq}%
  \BibitemOpen
  \bibfield{author}{%
  \bibinfo {author} {\bibfnamefont{W.}~\bibnamefont{Paul}}, \bibinfo {author}
  {\bibfnamefont{D.~Y.}\ \bibnamefont{Yoon}},\ and\ \bibinfo {author}
  {\bibfnamefont{G.~D.}\ \bibnamefont{Smith}},\ }%
  \bibfield{journal}{%
  \bibinfo {journal} {Journal of Chemical Physics}\ }%
  \textbf{\bibinfo {volume} {103}},\ \bibinfo {pages} {1702} (\bibinfo {year}
  {1995})\BibitemShut{NoStop}%
\bibitem{Waheed:2005km}%
  \BibitemOpen
  \bibfield{author}{%
  \bibinfo {author} {\bibfnamefont{N.}~\bibnamefont{Waheed}}, \bibinfo {author}
  {\bibfnamefont{M.~J.}\ \bibnamefont{Ko}},\ and\ \bibinfo {author}
  {\bibfnamefont{G.~C.}\ \bibnamefont{Rutledge}},\ }%
  \bibfield{journal}{%
  \bibinfo {journal} {Polymer}\ }%
  \textbf{\bibinfo {volume} {46}},\ \bibinfo {pages} {8689} (\bibinfo {year}
  {2005})\BibitemShut{NoStop}%
\bibitem{Limbach:2006cj}%
  \BibitemOpen
  \bibfield{author}{%
  \bibinfo {author} {\bibfnamefont{H.~J.}\ \bibnamefont{Limbach}}, \bibinfo
  {author} {\bibfnamefont{A.}~\bibnamefont{Arnold}}, \bibinfo {author}
  {\bibfnamefont{B.~A.}\ \bibnamefont{Mann}},\ and\ \bibinfo {author}
  {\bibfnamefont{C.}~\bibnamefont{Holm}},\ }%
  \bibfield{journal}{%
  \bibinfo {journal} {Computer Physics Communications}\ }%
  \textbf{\bibinfo {volume} {174}},\ \bibinfo {pages} {704} (\bibinfo {year}
  {2006})\BibitemShut{NoStop}%
\bibitem{Kolb:1999vt}%
  \BibitemOpen
  \bibfield{author}{%
  \bibinfo {author} {\bibfnamefont{A.}~\bibnamefont{Kolb}}\ and\ \bibinfo
  {author} {\bibfnamefont{B.}~\bibnamefont{D{\"u}nweg}},\ }%
  \bibfield{journal}{%
  \bibinfo {journal} {Journal of Chemical Physics}\ }%
  \textbf{\bibinfo {volume} {111}},\ \bibinfo {pages} {4453} (\bibinfo {year}
  {1999})\BibitemShut{NoStop}%
\bibitem{P.G.deGennes1995}%
  \BibitemOpen
  \bibfield{author}{%
  \bibinfo {author} {\bibfnamefont{P.~G.~de Gennes}\ \ and\ \bibnamefont{J.~Prost}},\ }%
  \emph{\bibinfo {title} {The Physics of Liquid Crystals}}\ (\bibinfo
  {publisher} {Oxford University Press},\ \bibinfo {year}
  {1995})\BibitemShut{NoStop}%
\bibitem{1983PhRvB..28..784S}%
  \BibitemOpen
  \bibfield{author}{%
  \bibinfo {author} {\bibfnamefont{P.~J.}\ \bibnamefont{Steinhardt}}, \bibinfo
  {author} {\bibfnamefont{D.~R.}\ \bibnamefont{Nelson}},\ and\ \bibinfo
  {author} {\bibfnamefont{M.}~\bibnamefont{Ronchetti}},\ }%
  \bibfield{journal}{%
  \bibinfo {journal} {Physical Review B}\ }%
  \textbf{\bibinfo {volume} {28}},\ \bibinfo {pages} {784} (\bibinfo {year}
  {1983})\BibitemShut{NoStop}%
\bibitem{Lechner:2008vz}%
  \BibitemOpen
  \bibfield{author}{%
  \bibinfo {author} {\bibfnamefont{W.}~\bibnamefont{Lechner}}\ and\ \bibinfo
  {author} {\bibfnamefont{C.}~\bibnamefont{Dellago}},\ }%
  \bibfield{journal}{%
  \bibinfo {journal} {Journal of Chemical Physics}\ }%
  \textbf{\bibinfo {volume} {129}},\ \bibinfo {pages} {114707} (\bibinfo {year}
  {2008})\BibitemShut{NoStop}%
\bibitem{Rycroft:2009ff}%
  \BibitemOpen
  \bibfield{author}{%
  \bibinfo {author} {\bibfnamefont{C.~H.}\ \bibnamefont{Rycroft}},\ }%
  \bibfield{journal}{%
  \bibinfo {journal} {Chaos}\ }%
  \textbf{\bibinfo {volume} {19}},\ \bibinfo {pages} {041111} (\bibinfo {year}
  {2009})\BibitemShut{NoStop}%
\bibitem{2002ARPC...53..291B}%
  \BibitemOpen
  \bibfield{author}{%
  \bibinfo {author} {\bibfnamefont{P.~G.}\ \bibnamefont{Bolhuis}}, \bibinfo
  {author} {\bibfnamefont{D.~D.}\ \bibnamefont{Chandler}}, \bibinfo {author}
  {\bibfnamefont{C.}~\bibnamefont{Dellago}},\ and\ \bibinfo {author}
  {\bibfnamefont{P.~L.}\ \bibnamefont{Geissler}},\ }%
  \bibfield{journal}{%
  \bibinfo {journal} {Annual Review of Physical Chemistry}\ }%
  \textbf{\bibinfo {volume} {53}},\ \bibinfo {pages} {291} (\bibinfo {year}
  {2002})\BibitemShut{NoStop}%
\bibitem{PhysRevE.63.061807}%
  \BibitemOpen
  \bibfield{author}{%
  \bibinfo {author} {\bibfnamefont{T.}~\bibnamefont{Miura}}, \bibinfo {author}
  {\bibfnamefont{R.}~\bibnamefont{Kishi}}, \bibinfo {author}
  {\bibfnamefont{M.}~\bibnamefont{Mikami}},\ and\ \bibinfo {author}
  {\bibfnamefont{Y.}~\bibnamefont{Tanabe}},\ }%
  \bibfield{journal}{%
  \Doi{10.1103/PhysRevE.63.061807}{\bibinfo {journal} {Phys. Rev. E}}\ }%
  \textbf{\bibinfo {volume} {63}},\ \bibinfo {pages} {061807} (\bibinfo {month}
  {May}\ \bibinfo {year} {2001})\BibitemShut{NoStop}%
\end{thebibliography}
\end{document}